\documentclass[intlimits,twoside,a4paper]{article}

\usepackage[cp1251]{inputenc}

\usepackage[eqsecnum]{cmpj3}

\articletype{A part of the Special collection to the 100th anniversary of the birth of Ihor Yukhnovskii}

\issue{2025}{28}{2}{23302}
\doinumber{10.5488/CMP.28.23302}

\title[Method of canonical transformations in the theory of quantum gases interacting with radiation]%
{Method of canonical transformations in the theory of quantum gases interacting with radiation%
}
\author[M. S. Bulakhov, A.~S.~Peletminskii, P.~P.~Kostrobij, I.~A.~Ryzha, Yu.~V.~Slyusarenko]{M.~S.~Bulakhov\orcid{0000-0002-8409-5558}\refaddr{label1},
	A.~S.~Peletminskii\orcid{0000-0001-6352-4838}\refaddr{label1},
	P.~P.~Kostrobij\orcid{0000-0002-4428-1647}\refaddr{label2},
	I.~A.~Ryzha\orcid{0000-0001-6647-0521}\refaddr{label2},
	Yu.~V.~Slyusarenko\orcid{0000-0001-5298-0731}\refaddr{label1,label2,label3}\thanks{Corresponding author: \email{slusarenko@kipt.kharkov.ua}.}
}
\addresses{
	\addr{label1} Akhiezer Institute for Theoretical Physics, NSC Kharkiv Institute of Physics and Technology, 1 Akademichna St., 61108 Kharkiv, Ukraine
	\addr{label2} Lviv Polytechnic National University, 12 S. Bandera St., 79013 Lviv, Ukraine
	\addr{label3} V.~N.~Karazin National University of Kharkiv, 4 Svobody Sq., 61022 Kharkiv,  Ukraine
}


\date{Received March 31, 2025, in final form May 25, 2025}

\begin{document}
	
	\maketitle
	
	\begin{abstract}
		An approach to the theoretical study of effects and phenomena in quantum gases interacting with radiation is proposed. The approach is based on a modification of the canonical transformation method, which was once used to diagonalize Hamiltonians describing the interaction of electrons with phonons in a solid. The capabilities of the method are demonstrated by studying the influence of photons on the spectral characteristics of atoms of quantum gases interacting with radiation. Within the framework of the developed approach, the effect of “dressing” atoms of quantum gases by a cloud of virtual photons is investigated and expressions for the energy characteristics of such dressed atoms --- quasiparticles are obtained. The problem of defining the concept of the effective mass of such quasiparticles is discussed.
		\printkeywords
	\end{abstract}

	\section*{Introduction}
	
	The problems of studying the influence of photons on processes and phenomena in low-temperature (quantum) gases, as well as the influence of gases on photons, can be considered, bearing in mind several different aspects. For example, from the point of view of photons that gain mass in low-temperature gases due to the interaction of radiation (photons) with such an environment. It is the influence of the medium that can lead to a significant change in the dispersion law of photons in the medium, being responsible for the appearance of the so-called ``cut-off frequency'' in it. In turn, the cut-off frequency forms the effective mass of the photon \cite{1,2,3}. Here, we should immediately emphasize that we are talking about the effective mass of the photon, since already in the medium the photon should be preferably considered as a quasiparticle, something similar to a polariton \cite{3,4}.
	
	This phenomenon of acquiring an effective mass in matter is associated with the capability of realizing in the medium such a phenomenon as Bose-Einstein condensation (BEC) of photons (an exotic phenomenon from the usual point of view of BEC in a gas of bosons), see \cite{1,2,3,4,5,6}. In a ``pure'' gas of photons (i.e., without a medium), such a phenomenon is generally impossible due to the absence of photon mass and the impossibility of thermalizing the gas of photons, since the cross sections of photon--photon scattering are extremely small \cite{7,8}. As we can see, the first obstacle is removed precisely due to the acquisition of the effective mass by a photon. Thermalization of photons takes place due to thermalization of the medium, wherein equilibrium is established after a change in temperature.
	
	Another aspect of the mutual influence of the photonic component and the low-temperature gas is related to the magnetic properties of the medium --- a quantum gas in equilibrium with radiation. As it turns out, the presence of the photonic component can significantly affect the magnetic properties of such a medium \cite{9}. Under certain conditions, the influence of photons on the system can lead to both an increase and a decrease in the magnetization of the system. Moreover, such an influence is not obvious a priori, since it is believed that the photon does not have a magnetic moment. We should note here that one can speak with certainty about the absence of a magnetic moment only regarding a photon in a vacuum, where, by the way, it does not have a mass. The question of  conditions at which a photon can acquire a magnetic moment and whether this can occur in the medium remains open at the moment \cite{10,11,12}. As in the two previous aspects (consequences) of the mutual influence of the photonic and atomic components, in this case, the fact of the existence of an effective mass of photons (or the acquisition of a cut-off frequency in the dispersion law) in the medium also plays an important role.
	
	However, in this connection, a question of a somewhat opposite plan also arises --- about the significance of the influence of the photonic component on the dispersion law of quantum gases. That is, to what extent is the dispersion law of atoms of quantum gases deformed (and how much does their mass change) due to their interaction with radiation --- photons? As is known, such interaction of radiation with the environment can be reduced to two types. Namely: processes of absorption-emission of photons by atoms, and processes of scattering of photons by atoms, see, for example, \cite{13,14}. In this case, the quantum mechanical states of atoms also change. It cannot be stated in advance that the influence of photons on the quantum mechanical states of atoms is unambiguously small. For example, the following argument can be given in support of the previous thesis. Below, in section~\ref{Sec1}, we give expressions for the Hamiltonian of a hydrogen-like plasma in an external electromagnetic field in the presence of radiation. It contains as a term an expression for the interaction responsible for the processes of absorption-emission of photons by atoms (see below, section~\ref{Sec1}). This term attracts attention by its analogy with the Fr\"ohlich Hamiltonian \cite{15}, which describes the interaction of electrons with phonons in a solid. As is known, such interaction leads to the appearance of attractive forces between electrons and the formation of Cooper pairs --- bosons, and thus becomes the cause of the phenomenon of superconductivity. In addition to a significant change in the forces of interaction between electrons (from repulsion to attraction), the influence of phonons can also lead to a significant change in the law of electron dispersion, which is usually interpreted as acquiring an effective mass by electrons (see, for example, \cite{16, 17}). The value of this mass can reach a significant value, comparable to the mass of a free electron \cite{18}.
	
	According to the above, there is a temptation to consider the problem of how the spectral characteristics of atoms of quantum gases are affected by their interaction with radiation (the photonic component of the system). In this article, the solution to such a problem is rather illustrative, ``without specificities'', that is, without ``bringing us to a number''. The main goal of this paper is to develop an approach to a description of the influence of radiation on the spectral characteristics of atoms outside the framework of perturbation theory and to demonstrate its capabilities from this point of view. The authors propose to construct a description of the mentioned influence using the method of canonical transformations, similar to the transformations formulated in \cite{16}. The method of canonical transformations \cite{16} for the problems announced here must be significantly modified. As we will see from the following presentation, the approach developed in the article allows us to obtain not only qualitative results illustrating the influence of the interaction of quantum gas atoms with radiation on spectral characteristics. We can also speak about specifying the tasks to achieve numerical results.
	
	\section{Hamiltonian of a low-temperature gas of hydrogen-like atoms interacting with radiation}\label{Sec1}
	
	Therefore, we solve the problem announced in the article, based on the above-mentioned analogy between the expression for the interaction responsible for the processes of absorption-emission of photons by atoms, and the Hamiltonian of the Fr\"ohlich type \cite{15}, which describes the interaction of electrons with phonons in a solid. However, let us first demonstrate this analogy. We will proceed from the Hamiltonian of a hydrogen-like weakly ionized plasma in an external electromagnetic field, which also takes into account the interaction of all components with radiation, that is, photons.
	
	Such a Hamiltonian is constructed in \cite{14}. In this paper, an approximate method of secondary quantization is proposed to describe multiparticle systems in the presence of bound particle states. For this purpose, a simple model system is considered, consisting of three different gas components: subsystems of two different oppositely charged fermions and their bound states --- atoms. In \cite{3, 14}, it is explained that the developed approximate method of secondary quantization is the most correct to be applied in cases of low temperatures. The reason is that the formulations of \cite{14} are valid if the average kinetic energy of the system particles is small compared to the energies of bound states (atoms). We should note here that the formulations of the proposed method of secondary quantization are successfully used in \cite{19} to describe the states of low-temperature gases of Fermi atoms of two different types in thermodynamic equilibrium with a gas of heteronuclear molecules formed by these fermions. This circumstance encourages us to use Hamiltonians \cite{14} to solve the problems of this paper.
	
	Let us again emphasize that in \cite{14} the complete Hamiltonian of a hydrogen-like weakly ionized plasma in an external electromagnetic field is constructed. This Hamiltonian takes into account the interaction of all components of this system with each other: the interaction of atoms with atoms, atoms with electrons and ions, ions with each other, electrons with each other, photons with all other components, and the interaction of all components with the electromagnetic field. However, as already noted above, we are interested in the effect of photons on the energy characteristics of atoms of quantum gases. Therefore, from the complete Hamiltonian, we consider here only that part of it that concerns free atoms and photons, as well as their interaction with each other. We  assume that the external electromagnetic field is absent. The easiest way to do this is to use the expression for the complete Hamiltonian in the form given in Section 3 of \cite{20}. We should emphasize that in \cite{20}, to simplify the calculations, the presence of spin variables as individual quantum characteristics of subsystems’ particles is not taken into account. It is not taken into account in this paper for the same reason. Moreover, the calculations necessary to obtain the results of this article are even more complicated and cumbersome.
	
	For bound states (hydrogen-like atoms --- bosons) with mass $M\equiv {{m}_{1}}+{{m}_{2}}$ and momentum $\mathbf{p}$, in the paper \cite{14} in the low-energy region, the possibility of introducing the creation $\hat{\eta }_{\alpha }^{+}\left( \mathbf{p} \right)$ and annihilation ${{\hat{\eta }}_{\alpha }}\left( \mathbf{p} \right)$ operators with the usual Bose commutation relations is proved:
	\begin{equation}\label{eq1}
		\left[ {{{\hat{\eta }}}_{\alpha }}\left( \mathbf{p} \right),\hat{\eta }_{\beta }^{+}\left( {\mathbf{{p}'}} \right) \right]\equiv {{\hat{\eta }}_{\alpha }}\left( \mathbf{p} \right)\hat{\eta }_{\beta }^{+}\left( {\mathbf{{p}'}} \right)-\hat{\eta }_{\beta }^{+}\left( {\mathbf{{p}'}} \right){{\hat{\eta }}_{\alpha }}\left( \mathbf{p} \right)=\Delta \left( \mathbf{p}-\mathbf{{p}'} \right){{\delta }_{\alpha \beta }},
	\end{equation}
	where index ``$\alpha $'' (or ``$\beta $'') denotes a set of quantum numbers characterizing the quantum-mechanical state of the hydrogen-like atom, values ${{\delta }_{\alpha \beta }}$ and $\Delta \left( \mathbf{p}-\mathbf{{p}'} \right)$ are the usual Kronecker symbols. For ease of calculations, the Kronecker symbol, whose arguments are vectors, is denoted by the uppercase letter ``delta'' of the Greek alphabet. As for the mass of the atom, the expression for it  $M\equiv {{m}_{1}}+{{m}_{2}}$ reflects the fact that the hydrogen-like atom is a bound state of an electron with mass ${{m}_{1}}$  and a core  ${{m}_{2}}$, for more details see \cite{14}.
	
	We also take into account the presence of photons in the system with the dispersion law  $\omega \left( k \right)$ ($\omega $  is the frequency, $\mathbf{k}$  is the wave vector), photon creation operators  $\hat{C}_{\lambda }^{+}\left( \mathbf{k} \right)$ with wave vector $\mathbf{k}$  and polarization   $\lambda =1,2$, and annihilation operators  ${{\hat{C}}_{\lambda }}\left( \mathbf{k} \right)$,
	\begin{equation}\label{eq2}
		\omega \left( k \right)=ck,	\quad \left[ {{{\hat{C}}}_{\lambda }}\left( \mathbf{k} \right),\hat{C}_{{{\lambda }'}}^{+}\left( {\mathbf{{k}'}} \right) \right]=\Delta \left( \mathbf{k}-\mathbf{{k}'} \right).
	\end{equation}
	All other components of the system, and therefore the corresponding parts of the Hamiltonian, as noted above, are not taken into account. Therefore, in terms of the introduced particle creation and annihilation operators, the desired Hamiltonian of a low-temperature gas of hydrogen-like atoms interacting with radiation, based on \cite{20}, can be written in the following form:
	\begin{equation}\label{eq3}
		\mathcal{\hat{H}}={{\mathcal{\hat{H}}}_{0}}+\mathcal{\hat{V}},
	\end{equation}
	where  $ {{\mathcal{\hat{H}}}_{0}}$ is the Hamiltonian of free particles:
	\begin{equation}\label{eq4}
		{{\mathcal{\hat{H}}}_{0}}=\sum\limits_{\alpha \mathbf{p}}{{{\varepsilon }_{\alpha }}\left( \mathbf{p} \right)}\hat{\eta }_{\alpha }^{+}\left( \mathbf{p} \right){{\hat{\eta }}_{\alpha }}\left( \mathbf{p} \right)+\sum\limits_{\mathbf{k},\lambda }{\omega \left( \mathbf{k} \right)\hat{C}_{\lambda }^{+}\left( \mathbf{k} \right){{{\hat{C}}}_{\lambda }}\left( \mathbf{k} \right)}, \quad	{{\varepsilon }_{\alpha }}\left( \mathbf{p} \right)={{\varepsilon }_{\alpha }}+\frac{{{\mathbf{p}}^{2}}}{2M},	\quad {{\omega }_{\mathbf{k}}}=c\left| \mathbf{k} \right|,
	\end{equation}
	and the value  ${\varepsilon }_{\alpha }<0$ is the energy of the bound state (or the characteristic of the energy level of a hydrogen-like atom) with a set of quantum numbers  $\alpha $. In \eqref{eq3}, \eqref{eq4} and subsequent calculations, as is usually done, we formally set Planck constant  $\hbar $ equal to one,  $\hbar \equiv 1$; if necessary, the dependence of the results on  $\hbar $ can be easily restored.
	
	We write the Hamiltonian of interaction $\mathcal{\hat{V}}$  of the system of atoms with photons in the form:
	\begin{equation}\label{eq5}
		\mathcal{\hat{V}}={{\mathcal{\hat{V}}}_{1}}+{{\mathcal{\hat{V}}}_{2}},
	\end{equation}
	where operators ${{\mathcal{\hat{V}}}_{1}}$, $ \ {{\mathcal{\hat{V}}}_{2}}$  are given by the expressions:
	\begin{equation}\label{eq6}
		{{\mathcal{\hat{V}}}_{1}}\equiv -\frac{1}{c}\int{\rd\mathbf{x}\,\mathbf{\hat{a}}\left( \mathbf{x} \right)}\mathbf{\hat{j}}\left( \mathbf{x} \right), \quad	{{\mathcal{\hat{V}}}_{2}}\equiv \frac{1}{2{{c}^{2}}}\int{\rd\mathbf{x}\,{{{\mathbf{\hat{a}}}}^{2}}\left( \mathbf{x} \right)\hat{I}\left( \mathbf{x} \right)},
	\end{equation}
	in which the operator of vector potential $\mathbf{\hat{a}}\left( \mathbf{x} \right)$  is introduced in terms of photon creation-annihilation operators (see, for example, \cite{21})
	\begin{equation}\label{eq7}
		\mathbf{\hat{a}}\left( \mathbf{x} \right)={{\left( \frac{2\piup }{V} \right)}^{{1}/{2}\;}}c\sum\limits_{\mathbf{k}}{\sum\limits_{\lambda =1}^{2}{{{\omega }^{-{1}/{2}\;}}\left( \mathbf{k} \right){{\mathbf{e}}_{\lambda }}\left( \mathbf{k} \right)\left\{ \hat{C}_{\lambda }^{+}\left( \mathbf{k} \right){{\re}^{-\ri\mathbf{kx}}}+{{{\hat{C}}}_{\lambda }}\left( \mathbf{k} \right){{\re}^{\ri\mathbf{kx}}} \right\}}}.
	\end{equation}
	In the last formula, $c$  is the speed of light in vacuum, $V$  is the volume of the system, ${\mathbf{e}}_{\mathbf{k}}^{(\lambda)} $  is the polarization vector of the photon in the state   $\mathbf{k}$ and $\lambda =1,2$. Note that Coulomb gauge is chosen for the radiation field.
	
	The current density operator $\mathbf{\hat{j}}\left( \mathbf{x} \right)$  in \eqref{eq6} is defined by the formulas:
	\begin{equation}\label{eq8}
		\mathbf{\hat{j}}\left( \mathbf{x} \right)=\frac{1}{V}\sum\limits_{\mathbf{p},\mathbf{{p}'}}{\sum\limits_{\alpha ,\beta }{{{\re}^{\ri\mathbf{x}\left( \mathbf{{p}'}-\mathbf{p} \right)}}}\left[ \frac{\left( \mathbf{p}+\mathbf{{p}'} \right)}{2M}{{\sigma }_{\alpha \beta }}\left( \mathbf{p}-\mathbf{{p}'} \right)+{{\mathbf{j}}_{\alpha \beta }}\left( \mathbf{p}-\mathbf{{p}'} \right) \right]}\hat{\eta }_{\alpha }^{+}\left( \mathbf{p} \right){{\hat{\eta }}_{\beta }}\left( {\mathbf{{p}'}} \right),
	\end{equation}
	where the tensors  ${{\sigma }_{\alpha \beta }}\left( \mathbf{k} \right)$,  ${{\mathbf{j}}_{\alpha \beta }}\left( \mathbf{k} \right)$    are given by expressions
	\begin{align}
		{{\sigma }_{\alpha \beta }}\left( \mathbf{k} \right)&\equiv e\int{\rd\,\mathbf{y}\varphi _{\alpha }^{*}}\left( \mathbf{y} \right){{\varphi }_{\beta }}\left( \mathbf{y} \right)\left[ \exp \left( -\ri\frac{{{m}_{1}}}{M}\mathbf{ky} \right)-\exp \left( \ri\frac{{{m}_{2}}}{M}\mathbf{ky} \right) \right], \quad M\equiv m_{1}+m_{2},\nonumber\\
		{{\mathbf{j}}_{\alpha \beta }}\left( \mathbf{k} \right)&\equiv e\frac{\ri}{2}\int{\rd\mathbf{y}\left( \varphi _{\alpha }^{*}\left( \mathbf{y} \right)\frac{\partial {{\varphi }_{\beta }}\left( \mathbf{y} \right)}{\partial \mathbf{y}}-\frac{\partial \varphi _{\alpha }^{*}\left( \mathbf{y} \right)}{\partial \mathbf{y}}{{\varphi }_{\beta }}\left( \mathbf{y} \right) \right)}\nonumber\\
	&	\times\left[ \frac{1}{{{m}_{1}}}\exp \left( \ri\frac{{{m}_{2}}}{M}\mathbf{ky} \right)
		+\frac{1}{{{m}_{2}}}\exp \left( -\ri\frac{{{m}_{1}}}{M}\mathbf{ky} \right) \right],
		\label{eq9}
	\end{align}
	where $M$  is the mass of an atom, ${{m}_{1}}$, ${{m}_{2}}$   are the masses of the electron and the core, respectively, and ${{\varphi }_{\alpha }}\left( \mathbf{y} \right)$   is the wave function of a hydrogen-like atom in a state $\alpha $   in coordinate space, which is considered to be known. In \cite{14, 20}, the wave functions  ${{\varphi }_{\alpha }}\left( \mathbf{y} \right)$  (and hence the sets of quantum mechanical characteristics of the atom  $\alpha $) refer to the discrete spectrum of a hydrogen-like atom with the completeness condition 
	\begin{equation}\label{eq10}
		\sum\limits_{\alpha }{\varphi _{\alpha }^{*}\left( \mathbf{y} \right){{\varphi }_{\alpha }}\left( \mathbf{x} \right)}=\delta \left( \mathbf{x}-\mathbf{y} \right),
	\end{equation}
	where  $\delta \left( \mathbf{x}-\mathbf{y} \right)$ is the Dirac delta function.
	
	Finally, the operator  $\hat{I}\left( \mathbf{x} \right)$, contained in \eqref{eq6}, according to \cite{14, 20}, can be written in the form
	\begin{equation}\label{eq11}
		\hat{I}\left( \mathbf{x} \right)\equiv \frac{1}{V}\sum\limits_{\mathbf{p},\mathbf{{p}'}}{\sum\limits_{\alpha ,\beta }{{{\re}^{\ri\mathbf{x}\left( \mathbf{{p}'}-\mathbf{p} \right)}}}}{{I}_{\alpha \beta }}\left( \mathbf{p}-\mathbf{{p}'} \right)\hat{\eta }_{\alpha }^{+}\left( \mathbf{p} \right){{\hat{\eta }}_{\beta }}\left( {\mathbf{{p}'}} \right),
	\end{equation}
	where the tensor ${{I}_{\alpha \beta }}\left( \mathbf{p}-\mathbf{{p}'} \right)$  is defined by expression:
	\begin{equation}\label{eq12}
		{{I}_{\alpha \beta }}\left( \mathbf{k} \right)\equiv {{e}^{2}}\int{\rd\,\mathbf{y}\varphi _{\alpha }^{*}}\left( \mathbf{y} \right){{\varphi }_{\beta }}\left( \mathbf{y} \right)\left[ \frac{1}{{{m}_{1}}}\exp \left( \ri\frac{{{m}_{2}}}{M}\mathbf{ky} \right)+\frac{1}{{{m}_{2}}}\exp \left( -\ri\frac{{{m}_{1}}}{M}\mathbf{ky} \right) \right].
	\end{equation}
	Thus, expressions \eqref{eq1}--\eqref{eq12} completely determine the Hamiltonian of a two-component system composed of quantum gases interacting with photons without taking into account the interaction of atoms with each other. It should be emphasized here that taking into account in this paper all the terms of the most complete Hamiltonian, which was obtained in \cite{14}, does not pose any fundamental difficulties. However, in reality, taking them into account, in addition to significantly complicating (or even making impossible) analytical calculations, can significantly complicate the illustrativeness of the results from the point of view of demonstrating the main goal of this paper.
	
	It is on the basis of the Hamiltonian defined by formulas \eqref{eq1}--\eqref{eq12} that we propose to solve the problem announced above about constructing an approach to describing the influence of the photonic component on the dispersion law of atoms of a quantum gas interacting with photons. For calculations, it is advisable to rewrite the mentioned Hamiltonian in a slightly different form. At the same time, the Hamiltonian ${{\mathcal{\hat{H}}}_{0}}$  will remain in the form in which it is presented in formula \eqref{eq4}. The changes will affect only the form of the interaction Hamiltonians  ${{\mathcal{\hat{V}}}_{1}}$, ${{\mathcal{\hat{V}}}_{2}}$, see \eqref{eq6}--\eqref{eq12}. The operator ${{\mathcal{\hat{V}}}_{1}}$   for further calculations will be presented in the following form:
	\begin{align}
		{{\mathcal{\hat{V}}}_{1}}&=\sum\limits_{\alpha ,\beta ,\lambda }\sum\limits_{\mathbf{p},\mathbf{k}}\left\{ {{F}_{\alpha ,\beta ,\lambda }}\left( \mathbf{p},\mathbf{k} \right)\hat{\eta }_{\alpha }^{+}\left( \mathbf{p} \right){{{\hat{\eta }}}_{\beta }}\left( \mathbf{p}-\mathbf{k} \right){{{\hat{C}}}_{\lambda }}\left( \mathbf{k} \right)\right. \nonumber\\
		&\left. + \,\,F_{\beta ,\alpha ,\lambda }^{*}\left( \mathbf{p},\mathbf{k} \right)\hat{\eta }_{\beta }^{+}\left( \mathbf{p}-\mathbf{k} \right)\hat{C}_{\lambda }^{+}\left( \mathbf{k} \right){{{\hat{\eta }}}_{\alpha }}\left( \mathbf{p} \right) \right\},
		\label{eq13}
	\end{align}
	where we introduce the value
	\begin{equation}\label{eq14}
		{{F}_{\alpha ,\beta ,\lambda }}\left( \mathbf{p},\mathbf{k} \right)\equiv -\frac{1}{c}\sqrt{\frac{2\piup }{\omega \left( \mathbf{k} \right)V}}\mathbf{e}_{\mathbf{k}}^{\left( \lambda  \right)}{{\mathbf{J}}_{\alpha ,\beta }}\left( \mathbf{p},\mathbf{k} \right),
	\end{equation}
	in which ${{\mathbf{J}}_{\alpha ,\beta }}\left( \mathbf{p},\mathbf{k} \right) $  is defined by expression
	\begin{equation}\label{eq15}
		{{\mathbf{J}}_{\alpha ,\beta }}\left( \mathbf{p},\mathbf{k} \right)\equiv \frac{2\mathbf{p}-\mathbf{k}}{M}{{\sigma }_{\alpha \beta }}\left( \mathbf{k} \right)+{{\mathbf{j}}_{\alpha \beta }}\left( \mathbf{k} \right)
	\end{equation}
	and satisfies, as is easy to verify from definitions \eqref{eq9}, the following condition:
	\begin{equation}\label{eq16}
		\mathbf{J}_{\beta ,\alpha }^{*}\left( \mathbf{p},\mathbf{k} \right)=\frac{2\mathbf{p}-\mathbf{k}}{M}{{\sigma }_{\beta \alpha }}\left( -\mathbf{k} \right)+{{\mathbf{j}}_{\beta \alpha }}\left( -\mathbf{k} \right).
	\end{equation}

	It is worth noting that in the ``point'' approximation for an atom, the expression for  ${{\mathbf{J}}_{\alpha ,\beta }}\left( \mathbf{p},\mathbf{k} \right) $ is much simplified, taking the form (see \cite{3, 14, 20}):
	\begin{equation}\label{eq17}
		{{\mathbf{J}}_{\alpha \beta }}\left( \mathbf{p},\mathbf{k} \right)\approx \ri\left\{ \left( {{\varepsilon }_{\alpha }}-{{\varepsilon }_{\beta }} \right){{\mathbf{d}}_{\alpha \beta }}+\frac{\mathbf{p}}{M}\left( \mathbf{k}{{\mathbf{d}}_{\alpha \beta }} \right) \right\}, 
	\end{equation}
	where the value  ${{\mathbf{d}}_{\alpha \beta }}$ is the tensor of dipole moments, which is determined according to \eqref{eq9} by formulas:
	\begin{equation}\label{eq18}
		{{\sigma }_{\alpha \beta }}\left( \mathbf{k} \right)\approx -\ri\mathbf{k}{{\mathbf{d}}_{\alpha \beta }},	\quad {{\mathbf{d}}_{\alpha \beta }}=e\int{\rd\,\mathbf{y}\varphi _{\alpha }^{*}}\left( \mathbf{y} \right)\mathbf{y}{{\varphi }_{\beta }}\left( \mathbf{y} \right).
	\end{equation}

	As for the interaction operator  ${{\mathcal{\hat{V}}}_{2}}$ (see \eqref{eq6}), it can be written in a more convenient form for the following purposes:
	\begin{align}
		{{\mathcal{\hat{V}}}_{2}}&=\sum\limits_{{{\lambda }_{1}},{{\lambda }_{2}},{{\mathbf{k}}_{1}},{{\mathbf{k}}_{2}}}{\sum\limits_{\mathbf{p},\mathbf{{p}'}}{\sum\limits_{\alpha ,\beta }{{{G}_{\alpha ,\beta ,{{\lambda }_{1}},{{\lambda }_{2}}}}\left( \mathbf{p}-\mathbf{{p}'},{{\mathbf{k}}_{1}},{{\mathbf{k}}_{2}} \right)\hat{\eta }_{\alpha }^{+}\left( \mathbf{p} \right){{{\hat{\eta }}}_{\beta }}\left( {\mathbf{{p}'}} \right)}}} \nonumber\\
		&\times \left\{ {{{\hat{C}}}_{{{\lambda }_{1}}}}\left( {{\mathbf{k}}_{1}} \right){{{\hat{C}}}_{{{\lambda }_{2}}}}\left( {{\mathbf{k}}_{2}} \right)\Delta \left( {{\mathbf{k}}_{1}}+{{\mathbf{k}}_{2}}+\mathbf{{p}'}-\mathbf{p} \right)+\hat{C}_{{{\lambda }_{1}}}^{+}\left( {{\mathbf{k}}_{1}} \right)\hat{C}_{{{\lambda }_{2}}}^{+}\left( {{\mathbf{k}}_{2}} \right)\Delta \left( {{\mathbf{k}}_{1}}+{{\mathbf{k}}_{2}}-\mathbf{{p}'}+\mathbf{p} \right) \right.\nonumber\\
		&+\left. {{{\hat{C}}}_{{{\lambda }_{1}}}}\left( {{\mathbf{k}}_{1}} \right)\hat{C}_{{{\lambda }_{2}}}^{+}\left( {{\mathbf{k}}_{2}} \right)\Delta \left( {{\mathbf{k}}_{1}}-{{\mathbf{k}}_{2}}+\mathbf{{p}'}-\mathbf{p} \right)+\hat{C}_{{{\lambda }_{1}}}^{+}\left( {{\mathbf{k}}_{1}} \right){{{\hat{C}}}_{{{\lambda }_{2}}}}\left( {{\mathbf{k}}_{2}} \right)\Delta \left( {{\mathbf{k}}_{1}}-{{\mathbf{k}}_{2}}-\mathbf{{p}'}+\mathbf{p} \right) \right\},
		\label{eq19}
	\end{align}
	where
	\begin{align}
		{{G}_{\alpha ,\beta ,{{\lambda }_{1}},{{\lambda }_{2}}}}\left( \mathbf{p}-\mathbf{{p}'},{{\mathbf{k}}_{1}},{{\mathbf{k}}_{2}} \right)&\equiv \frac{\piup }{V{{c}^{2}}}{{\left( \frac{1}{\omega \left( {{\mathbf{k}}_{1}} \right)\omega \left( {{\mathbf{k}}_{2}} \right)} \right)}^{1/2}}\mathbf{e}_{{{\mathbf{k}}_{1}}}^{\left( {{\lambda }_{1}} \right)}\mathbf{e}_{{{\mathbf{k}}_{2}}}^{\left( {{\lambda }_{2}} \right)}{{I}_{\alpha \beta }}\left( \mathbf{p}-\mathbf{{p}'} \right)\nonumber\\
		&={{G}_{\alpha ,\beta ,{{\lambda }_{2}},{{\lambda }_{1}}}}\left( \mathbf{p}-\mathbf{{p}'},{{\mathbf{k}}_{2}},{{\mathbf{k}}_{1}} \right).
		\label{eq20}
	\end{align}
	We should note here that the value of  ${{I}_{\alpha \beta }}\left( \mathbf{p}-\mathbf{{p}'} \right)$  in the main approximation of a ``point'' atom also takes on a simplified form 
	\begin{equation}\label{eq21}
		{{I}_{\alpha \beta }}\left( \mathbf{p}-\mathbf{{p}'} \right)\approx \frac{{{e}^{2}}}{{\tilde{m}}}{{\delta }_{\alpha \beta }}, \quad \tilde{m}\equiv \frac{{{m}_{1}}{{m}_{2}}}{{{m}_{1}}+{{m}_{2}}},
	\end{equation}
	where $\tilde{m}$  is the reduced mass.
	
	As we can see, the Hamiltonian  ${{\mathcal{\hat{V}}}_{1}}$, written in the form \eqref{eq13}--\eqref{eq18}, describes the processes of emission and absorption of photons by atoms, while the Hamiltonian ${{\mathcal{\hat{V}}}_{2}}$  is responsible for describing the processes associated with the scattering of photons by atoms \cite{14}. It is the Hamiltonian ${{\mathcal{\hat{V}}}_{1}}$  that attracts attention due to its similarity to the well-known Fr\"oelich Hamiltonian \cite{15}, which describes the processes of absorption and emission of phonons by electrons. In papers \cite{16, 17}, similar Hamiltonians are also used to describe some aspects of the interaction of electrons with phonons, in particular, to explain the mechanism of electrons acquiring an effective mass. In \cite{16}, a canonical transformation is proposed that eliminates the trilinear exciton-phonon interaction and introduces new elementary excitations --- ``dressed'' excitons. The Hamiltonian describing the mentioned exciton-phonon interaction is mathematically similar to the Hamiltonian  ${{\mathcal{\hat{V}}}_{1}}$, although the latter is much more complex. It should be noted that usually such canonical transformations can be used to diagonalize certain terms of the full Hamiltonian of interacting particles without using the methods of perturbation theory for weak interaction. In this case, the terms in the full Hamiltonian not involved in the diagonalization (as a rule, the residual interaction, \cite{15, 16}) are redefined in terms of the operators of creation-annihilation of new elementary excitations (quasiparticles). These new elementary excitations appear as a consequence of diagonalization. The redefined interaction (quasiparticle interaction) may experience significant changes in the properties compared to the original properties.
	
	Recall that the purpose of this paper is to analyze the influence of the interaction of quantum gases with radiation (the photonic component of the system) on the spectral characteristics of quantum gases. When solving this problem, it is necessary to transform the Hamiltonian ${{\mathcal{\hat{H}}}}={{\mathcal{\hat{H}}}_{0}}+{{\mathcal{\hat{V}}}_{1}}$ in such a way as to get rid of the three-operator interaction Hamiltonian  ${{\mathcal{\hat{V}}}_{1}}$. In other words, we must introduce quasiparticles in order to diagonalize the so-called truncated Hamiltonian
	\begin{equation}\label{eq22}
		{{\mathcal{\hat{H}}}_{\mathrm{tr}}}\left( \eta ,C \right)={{\mathcal{\hat{H}}}_{0}}\left( \eta ,C \right)+{{\mathcal{\hat{V}}}_{1}}\left( \eta ,C \right),
	\end{equation}
	where the Hamiltonian of free particles $ {{\mathcal{\hat{H}}}_{0}}\left( \eta ,C \right)$  is defined by formula \eqref{eq4}, and  ${{\mathcal{\hat{V}}}_{1}}\left( \eta ,C \right)$ is given by expressions \eqref{eq13}--\eqref{eq18}. For this purpose, we use the paper \cite{16}, modifying its method of canonical transformations to the specificity of the system of quantum gases interacting with radiation. Note that we do not take the Hamiltonian ${{\mathcal{\hat{V}}}_{2}}\left( \eta ,C \right)$ (see \eqref{eq19}--\eqref{eq21}) into account yet. 
	
	\section{Canonical transformations and operators of creation and annihilation of elementary excitations (quasiparticles)}\label{Sec2}
	
	By analogy with \cite{16}, we assume that new (non-decaying) elementary excitations can be obtained by diagonalizing the operator \eqref{eq22} under canonical transformation $\re^{S\left( \eta,C \right)}$
	\begin{equation}\label{eq23}
		{{\hat{\gamma }}_{\alpha }}\left( \mathbf{p} \right)={{\re}^{S\left( \eta,C \right)}}{{\hat{\eta }}_{\alpha }}\left( \mathbf{p} \right){{\re}^{-S\left( \eta,C \right)}},	\quad {{\hat{\chi }}_{\lambda }}\left( \mathbf{k} \right)={{\re}^{S\left( \eta,C \right)}}{{\hat{C}}_{\lambda }}\left( \mathbf{k} \right){{\re}^{-S\left( \eta,C \right)}},
	\end{equation}
	generated by the anti-Hermitian transformation
	\begin{align}
		S\left( \eta,C  \right)=\sum\limits_{\lambda ;\mathbf{k}}\sum\limits_{\alpha ,\beta ;\mathbf{p}}\left\{ {{\Phi }_{\alpha ,\beta ,\lambda }}\left( \mathbf{p},\mathbf{k} \right)\hat{\eta }_{\alpha }^{+}\left( \mathbf{p} \right){{{\hat{\eta }}}_{\beta }}\left( \mathbf{p}-\mathbf{k} \right){{{\hat{C }}}_{\lambda }}\left( \mathbf{k} \right)\right. \nonumber\\
		\left. -\,\,\Phi _{\beta ,\alpha ,\lambda }^{*}\left( \mathbf{p},\mathbf{k} \right)\hat{\eta }_{\beta }^{+}\left( \mathbf{p}-\mathbf{k} \right)\hat{C }_{\lambda }^{+}\left( \mathbf{k} \right){{{\hat{\eta }}}_{\alpha }}\left( \mathbf{p} \right) \right\}.
		\label{eq24}
	\end{align}
	The unknown functions ${{\Phi }_{\alpha ,\beta ,\lambda }}\left( \mathbf{p},\mathbf{k} \right)$ included in \eqref{eq24} must be defined in such a way that after the transition to the new operators  ${{\hat{\gamma }}_{\alpha }}\left( \mathbf{p} \right)$,  ${{\hat{\chi }}_{\lambda }}\left( \mathbf{k} \right)$ there will be no terms in the Hamiltonian that are linear in  ${{\hat{\chi }}_{\lambda }}\left( \mathbf{k} \right)$.
	
	To transform the Hamiltonian \eqref{eq22} to the new operators, we first introduce an auxiliary Hamiltonian~\cite{16}
	\begin{equation}\label{eq24-1}
		{{\mathcal{\hat{\tilde{H}}}}_{\mathrm{tr}}}\left( \gamma ,\chi  \right)={{\re}^{S\left( \eta,C \right)}}{{\mathcal{\hat{H}}}_{\mathrm{tr}}}\left( \eta ,C \right){{\re}^{-S\left( \eta,C \right)}}.
	\end{equation}
	Then, using \eqref{eq23} and taking into account \eqref{eq4}, \eqref{eq13}, we can arrive at the following expression
	\begin{align}
		&{{\mathcal{\hat{\tilde{H}}}}_{\mathrm{tr}}}\left( \gamma ,\chi  \right)=\sum\limits_{\alpha \mathbf{p}}{{{\varepsilon }_{\alpha }}\left( \mathbf{p} \right)}\hat{\gamma }_{\alpha }^{+}\left( \mathbf{p} \right){{\hat{\gamma }}_{\alpha }}\left( \mathbf{p} \right)+\sum\limits_{\mathbf{k},\lambda }{\omega \left( \mathbf{k} \right)\hat{\chi }_{\lambda }^{+}\left( \mathbf{k} \right){{{\hat{\chi }}}_{\lambda }}}\left( \mathbf{k} \right)\nonumber\\
	&	+\sum\limits_{\alpha ,\beta ,\lambda }{\sum\limits_{\mathbf{p},\mathbf{k}}{\left\{ {{F}_{\alpha ,\beta ,\lambda }}\left( \mathbf{p},\mathbf{k} \right)\hat{\gamma }_{\alpha }^{+}\left( \mathbf{p} \right){{{\hat{\gamma }}}_{\beta }}\left( \mathbf{p}-\mathbf{k} \right){{{\hat{\chi }}}_{\lambda }}\left( \mathbf{k} \right)+F_{\beta ,\alpha ,\lambda }^{*}\left( \mathbf{p},\mathbf{k} \right)\hat{\gamma }_{\beta }^{+}\left( \mathbf{p}-\mathbf{k} \right)\hat{\chi }_{\lambda }^{+}\left( \mathbf{k} \right){{{\hat{\gamma }}}_{\alpha }}\left( \mathbf{p} \right) \right\}}},
		\label{eq25}
	\end{align}
	where the values ${{F}_{\alpha ,\beta ,\lambda }}\left( \mathbf{p},\mathbf{k} \right)$  are defined by expressions \eqref{eq14}--\eqref{eq18}. Then the initial Hamiltonian~\eqref{eq22} in terms of the new operators  ${{\hat{\gamma }}_{\alpha }}\left( \mathbf{p} \right)$, ${{\hat{\chi }}_{\lambda }}\left( \mathbf{k} \right)$  is expressed using the following unitary transformation  ${{\re}^{S\left( \gamma ,\chi  \right)}}$  (cf. with \cite{16}):
	\begin{equation}\label{eq26}
		{{\mathcal{\hat{H}}}_{\mathrm{tr}}}\left( \gamma ,\chi  \right)={{\re}^{-S\left( \gamma ,\chi  \right)}}{{\mathcal{\hat{\tilde{H}}}}_{\mathrm{tr}}}\left( \gamma ,\chi  \right){{\re}^{S\left( \gamma ,\chi  \right)}},
	\end{equation}
	where [see \eqref{eq24}]
	\begin{align}
		S\left( \gamma ,\chi  \right)=\sum\limits_{\lambda ;\mathbf{k}}\sum\limits_{\alpha ,\beta ;\mathbf{p}}\left\{ {{\Phi }_{\alpha ,\beta ,\lambda }}\left( \mathbf{p},\mathbf{k} \right)\hat{\gamma }_{\alpha }^{+}\left( \mathbf{p} \right){{{\hat{\gamma }}}_{\beta }}\left( \mathbf{p}-\mathbf{k} \right){{{\hat{\chi }}}_{\lambda }}\left( \mathbf{k} \right)\right. \nonumber\\
		\left. -\,\,\Phi _{\beta ,\alpha ,\lambda }^{*}\left( \mathbf{p},\mathbf{k} \right)\hat{\gamma }_{\beta }^{+}\left( \mathbf{p}-\mathbf{k} \right)\hat{\chi }_{\lambda }^{+}\left( \mathbf{k} \right){{{\hat{\gamma }}}_{\alpha }}\left( \mathbf{p} \right) \right\}.
		\label{eq27}
	\end{align}
	In what follows, we are interested only in the eigenvalues of the Hamiltonian \eqref{eq26} corresponding to single-photon excitations, and therefore in ${{\mathcal{\hat{H}}}_{\mathrm{tr}}}\left( \gamma ,\chi  \right)$   we neglect the terms containing four or more operators  $\hat{\gamma }$, $\hat{\chi }$. The discarded operators must describe the residual interaction of new elementary excitations, which are determined by eigenfunctions of the type  $\hat{\gamma }_{\alpha \mathbf{p}}^{+}\left| 0 \right\rangle $, $\hat{\gamma }_{\alpha \mathbf{p}-\mathbf{k}}^{+}\hat{\chi }_{\mathbf{k}\lambda }^{+}\left| 0 \right\rangle $, where $\left| 0 \right\rangle $   is the state of absolute vacuum.
	
	Due to the unitary nature of the transformations \eqref{eq26}, the calculation of the operator ${{\mathcal{\hat{H}}}_{\mathrm{tr}}}\left( \gamma ,\chi  \right)$  must begin with finding the operators
	\begin{align}
		{{\re}^{-S\left( \gamma ,\chi  \right)}}{{\hat{\chi }}_{\lambda }}\left( \mathbf{k} \right){{\re}^{S\left( \gamma ,\chi  \right)}}, \quad {{\re}^{-S\left( \gamma ,\chi  \right)}}\hat{\chi }_{\lambda }^{+}\left( \mathbf{k} \right){{\re}^{S\left( \gamma ,\chi  \right)}}, \nonumber\\
		{{\re}^{-S\left( \gamma ,\chi  \right)}}{{\hat{\gamma }}_{\alpha }}\left( \mathbf{p} \right){{\re}^{S\left( \gamma ,\chi  \right)}}, \quad  {{\re}^{-S\left( \gamma ,\chi  \right)}}\hat{\gamma }_{\alpha }^{+}\left( \mathbf{p} \right){{\re}^{S\left( \gamma ,\chi  \right)}}.
		\label{eq28}
	\end{align}
	To calculate operators \eqref{eq28}, we need to use the general expression
	\begin{equation}\label{eq29}
		{{\re}^{-S}}A{{\re}^{S}}=A+\left[ A,S \right]+\frac{1}{2}\left[ \left[ A,S \right],S \right]+\dots,
	\end{equation}
	valid for an arbitrary operator  $A\left( \gamma ,\chi  \right)$. In this case, we must also keep in mind the general relations for the three commutators of arbitrary operators  $A$, $B$, $C$:
	\begin{equation}\label{eq30}
		\left[ A,BC \right]=\left[ A,B \right]C+B\left[ A,C \right]
	\end{equation}
	and the commutation rules \eqref{eq1}, \eqref{eq2}.
	
	Taking into account the last remarks, the results of calculations for the operator ${{\hat{\chi }}_{\lambda }}\left( \mathbf{k} \right)$  [see \eqref{eq28}] with an accuracy of a double commutator, as in expression \eqref{eq29}, can be represented in the form:
	\begin{align}
	&	{{\re}^{-S\left( \gamma ,\chi  \right)}}{{\hat{\chi }}_{\lambda }}\left( \mathbf{k} \right){{\re}^{S\left( \gamma ,\chi  \right)}}={{\hat{\chi }}_{\lambda }}\left( \mathbf{k} \right)-\sum\limits_{{{\alpha }_{1}},{{\beta }_{1}}}{\sum\limits_{{{\mathbf{p}}_{1}}}{\Phi _{{{\beta }_{1}},{{\alpha }_{1}},{{\lambda }_{1}}}^{*}\left( {{\mathbf{p}}_{1}},\mathbf{k} \right)\hat{\gamma }_{{{\beta }_{1}}}^{+}\left( {{\mathbf{p}}_{1}}-\mathbf{k} \right){{{\hat{\gamma }}}_{{{\alpha }_{1}}}}}}\left( {{\mathbf{p}}_{1}} \right)\nonumber\\
	&	-\frac{1}{2}\sum\limits_{\begin{smallmatrix} 
				{{\lambda }_{2}},{{\alpha }_{2}}, \\ 
				{{\alpha }_{1}},{{\beta }_{1}} 
		\end{smallmatrix}}{\sum\limits_{{{\mathbf{k}}_{2}},{{\mathbf{p}}_{1}}}{{{\Phi }_{{{\alpha }_{1}},{{\alpha }_{2}},{{\lambda }_{2}}}}\left( {{\mathbf{p}}_{1}},{{\mathbf{k}}_{2}} \right)}}\Phi _{{{\beta }_{1}},{{\alpha }_{1}},{{\lambda }_{1}}}^{*}\left( {{\mathbf{p}}_{1}},\mathbf{k} \right)\hat{\gamma }_{{{\beta }_{1}}}^{+}\left( {{\mathbf{p}}_{1}}-\mathbf{k} \right){{\hat{\gamma }}_{{{\alpha }_{2}}}}\left( {{\mathbf{p}}_{1}}-{{\mathbf{k}}_{2}} \right){{\hat{\chi }}_{{{\lambda }_{2}}}}\left( {{\mathbf{k}}_{2}} \right)\nonumber\\
	&	+\frac{1}{2}\sum\limits_{\begin{smallmatrix} 
				{{\lambda }_{2}},{{\alpha }_{2}}; \\ 
				{{\alpha }_{1}},{{\beta }_{1}} 
		\end{smallmatrix}}{\sum\limits_{{{\mathbf{k}}_{2}},{{\mathbf{p}}_{1}}}{{{\Phi }_{{{\alpha }_{2}},{{\beta }_{1}},{{\lambda }_{2}}}}\left( {{\mathbf{p}}_{1}}+{{\mathbf{k}}_{2}}-\mathbf{k},{{\mathbf{k}}_{2}} \right)\Phi _{{{\beta }_{1}},{{\alpha }_{1}},{{\lambda }_{1}}}^{*}\left( {{\mathbf{p}}_{1}},\mathbf{k} \right)\hat{\gamma }_{{{\alpha }_{2}}}^{+}\left( {{\mathbf{p}}_{1}}+{{\mathbf{k}}_{2}}-\mathbf{k} \right){{{\hat{\gamma }}}_{{{\alpha }_{1}}}}\left( {{\mathbf{p}}_{1}} \right){{{\hat{\chi }}}_{{{\lambda }_{2}}}}\left( {{\mathbf{k}}_{2}} \right)}}\nonumber\\
	&	+\frac{1}{2}\sum\limits_{\begin{smallmatrix} 
				{{\lambda }_{2}},{{\alpha }_{2}}; \\ 
				{{\alpha }_{1}},{{\beta }_{1}} 
		\end{smallmatrix}}{\sum\limits_{{{\mathbf{k}}_{2}},{{\mathbf{p}}_{1}}}{\Phi _{{{\alpha }_{1}},{{\alpha }_{2}},{{\lambda }_{2}}}^{*}\left( {{\mathbf{p}}_{1}}+{{\mathbf{k}}_{2}},{{\mathbf{k}}_{2}} \right)\Phi _{{{\beta }_{1}},{{\alpha }_{1}},{{\lambda }_{1}}}^{*}\left( {{\mathbf{p}}_{1}},\mathbf{k} \right)\hat{\chi }_{{{\lambda }_{2}}}^{+}\left( {{\mathbf{k}}_{2}} \right)\hat{\gamma }_{{{\beta }_{1}}}^{+}\left( {{\mathbf{p}}_{1}}-\mathbf{k} \right){{{\hat{\gamma }}}_{{{\alpha }_{2}}}}\left( {{\mathbf{p}}_{1}}+{{\mathbf{k}}_{2}} \right)}}\nonumber\\
		&-\frac{1}{2}\sum\limits_{\begin{smallmatrix} 
				{{\lambda }_{2}},{{\alpha }_{2}}; \\ 
				{{\alpha }_{1}},{{\beta }_{1}} 
		\end{smallmatrix}}{\sum\limits_{{{\mathbf{k}}_{2}},{{\mathbf{p}}_{1}}}{\Phi _{{{\alpha }_{2}},{{\beta }_{1}},{{\lambda }_{2}}}^{*}\left( {{\mathbf{p}}_{1}}-\mathbf{k},{{\mathbf{k}}_{2}} \right)\Phi _{{{\beta }_{1}},{{\alpha }_{1}},{{\lambda }_{1}}}^{*}\left( {{\mathbf{p}}_{1}},\mathbf{k} \right)\hat{\chi }_{{{\lambda }_{2}}}^{+}\left( {{\mathbf{k}}_{2}} \right)\hat{\gamma }_{{{\alpha }_{2}}}^{+}\left( {{\mathbf{p}}_{1}}-\mathbf{k}-{{\mathbf{k}}_{2}} \right)}}{{\hat{\gamma }}_{{{\alpha }_{1}}}}\left( {{\mathbf{p}}_{1}} \right)+\dots				
		\label{eq31}
	\end{align}
	We also present a similar result of transformations for the operator  ${{\hat{\gamma }}_{\alpha }}\left( \mathbf{p} \right)$:
	\begin{align}
	&	{{\re}^{-S\left( \gamma ,\chi  \right)}}{{\hat{\gamma }}_{\alpha }}\left( \mathbf{p} \right){{\re}^{S\left( \gamma ,\chi  \right)}}={{\hat{\gamma }}_{\alpha }}\left( \mathbf{p} \right)-\frac{1}{2}\sum\limits_{{{\lambda }_{1}},{{\alpha }_{1}},{{\alpha }_{2}}}{\sum\limits_{{{\mathbf{k}}_{1}}}{{{\Phi }_{\alpha ,{{\alpha }_{1}},{{\lambda }_{1}}}}\left( \mathbf{p},{{\mathbf{k}}_{1}} \right)\Phi _{{{\alpha }_{1}},{{\alpha }_{2}},{{\lambda }_{1}}}^{*}\left( \mathbf{p},{{\mathbf{k}}_{1}} \right){{{\hat{\gamma }}}_{{{\alpha }_{2}}}}}}\left( \mathbf{p} \right)\nonumber \\
		&+\sum\limits_{{{\lambda }_{1}},{{\alpha }_{1}}}\sum\limits_{{{\mathbf{k}}_{1}}}{{{\Phi }_{\alpha ,{{\alpha }_{1}},{{\lambda }_{1}}}}\left( \mathbf{p},{{\mathbf{k}}_{1}} \right){{{\hat{\gamma }}}_{{{\alpha }_{1}}}}\left( \mathbf{p}-{{\mathbf{k}}_{1}} \right){{{\hat{\chi }}}_{{{\lambda }_{1}}}}\left( {{\mathbf{k}}_{1}} \right)}\nonumber\\
	&	-\sum\limits_{{{\lambda }_{1}},{{\alpha }_{1}}}{\sum\limits_{{{\mathbf{k}}_{1}}}{\Phi _{\alpha ,{{\alpha }_{1}},{{\lambda }_{1}}}^{*}\left( {{\mathbf{k}}_{1}}+\mathbf{p},{{\mathbf{k}}_{1}} \right)\hat{\chi }_{{{\lambda }_{1}}}^{+}\left( {{\mathbf{k}}_{1}} \right){{{\hat{\gamma }}}_{{{\alpha }_{1}}}}}}\left( {{\mathbf{k}}_{1}}+\mathbf{p} \right)\nonumber\\
	&	+\frac{1}{2}\sum\limits_{\begin{smallmatrix} 
				{{\lambda }_{1}},{{\lambda }_{2}}, \\ 
				{{\alpha }_{1}},{{\alpha }_{2}} 
		\end{smallmatrix}}{\sum\limits_{{{\mathbf{k}}_{1}},{{\mathbf{k}}_{2}}}{{{\Phi }_{\alpha ,{{\alpha }_{1}},{{\lambda }_{1}}}}\left( \mathbf{p},{{\mathbf{k}}_{1}} \right){{\Phi }_{{{\alpha }_{1}},{{\alpha }_{2}},{{\lambda }_{2}}}}\left( \mathbf{p}-{{\mathbf{k}}_{1}},{{\mathbf{k}}_{2}} \right){{{\hat{\gamma }}}_{{{\alpha }_{2}}}}\left( \mathbf{p}-{{\mathbf{k}}_{1}}-{{\mathbf{k}}_{2}} \right){{{\hat{\chi }}}_{{{\lambda }_{2}}}}\left( {{\mathbf{k}}_{2}} \right){{{\hat{\chi }}}_{{{\lambda }_{1}}}}}}\left( {{\mathbf{k}}_{1}} \right)\nonumber\\
		&-\frac{1}{2}\sum\limits_{\begin{smallmatrix} 
				{{\lambda }_{1}},{{\alpha }_{1}}, \\ 
				{{\lambda }_{2}},{{\alpha }_{2}} 
		\end{smallmatrix}}{\sum\limits_{{{\mathbf{k}}_{2}},{{\mathbf{k}}_{1}}}{{{\Phi }_{\alpha ,{{\alpha }_{1}},{{\lambda }_{1}}}}\left( \mathbf{p},{{\mathbf{k}}_{1}} \right)\Phi _{{{\alpha }_{1}},{{\alpha }_{2}},{{\lambda }_{2}}}^{*}\left( \mathbf{p}-{{\mathbf{k}}_{1}}+{{\mathbf{k}}_{2}},{{\mathbf{k}}_{2}} \right)\hat{\chi }_{{{\lambda }_{2}}}^{+}\left( {{\mathbf{k}}_{2}} \right){{{\hat{\chi }}}_{{{\lambda }_{1}}}}\left( {{\mathbf{k}}_{1}} \right){{{\hat{\gamma }}}_{{{\alpha }_{2}}}}\left( \mathbf{p}-{{\mathbf{k}}_{1}}+{{\mathbf{k}}_{2}} \right)}}\nonumber\\
		&-\frac{1}{2}\sum\limits_{\begin{smallmatrix} 
				{{\lambda }_{1}},{{\alpha }_{1}}, \\ 
				{{\alpha }_{2}},{{\beta }_{2}} 
		\end{smallmatrix}}{\sum\limits_{{{\mathbf{p}}_{2}},{{\mathbf{k}}_{1}}}{{{\Phi }_{\alpha ,{{\alpha }_{1}},{{\lambda }_{1}}}}\left( \mathbf{p},{{\mathbf{k}}_{1}} \right)\Phi _{{{\beta }_{2}},{{\alpha }_{2}},{{\lambda }_{1}}}^{*}\left( {{\mathbf{p}}_{2}},{{\mathbf{k}}_{1}} \right)\hat{\gamma }_{{{\beta }_{2}}}^{+}\left( {{\mathbf{p}}_{2}}-{{\mathbf{k}}_{1}} \right){{{\hat{\gamma }}}_{{{\alpha }_{1}}}}\left( \mathbf{p}-{{\mathbf{k}}_{1}} \right){{{\hat{\gamma }}}_{{{\alpha }_{2}}}}}}\left( {{\mathbf{p}}_{2}} \right)\nonumber\\
	&	-\frac{1}{2}\sum\limits_{\begin{smallmatrix} 
				{{\lambda }_{1}},{{\alpha }_{1}}, \\ 
				{{\lambda }_{2}},{{\alpha }_{2}} 
		\end{smallmatrix}}{\sum\limits_{{{\mathbf{k}}_{1}},{{\mathbf{k}}_{2}}}{\Phi _{\alpha ,{{\alpha }_{1}},{{\lambda }_{1}}}^{*}\left( {{\mathbf{k}}_{1}}+\mathbf{p},{{\mathbf{k}}_{1}} \right){{\Phi }_{{{\alpha }_{1}},{{\alpha }_{2}},{{\lambda }_{2}}}}\left( {{\mathbf{k}}_{1}}+\mathbf{p},{{\mathbf{k}}_{2}} \right)\hat{\chi }_{{{\lambda }_{1}}}^{+}\left( {{\mathbf{k}}_{1}} \right){{{\hat{\chi }}}_{{{\lambda }_{2}}}}\left( {{\mathbf{k}}_{2}} \right){{{\hat{\gamma }}}_{{{\alpha }_{2}}}}\left( {{\mathbf{k}}_{1}}+\mathbf{p}-{{\mathbf{k}}_{2}} \right)}}\nonumber\\
	&	+\frac{1}{2}\sum\limits_{\begin{smallmatrix} 
				{{\lambda }_{1}},{{\lambda }_{2}},{{\alpha }_{1}}, \\ 
				{{\alpha }_{2}},{{\beta }_{2}} 
		\end{smallmatrix}}{\sum\limits_{{{\mathbf{p}}_{2}},{{\mathbf{k}}_{1}}}{\Phi _{\alpha ,{{\alpha }_{1}},{{\lambda }_{1}}}^{*}\left( {{\mathbf{k}}_{1}}+\mathbf{p},{{\mathbf{k}}_{1}} \right){{\Phi }_{{{\alpha }_{2}},{{\beta }_{2}},{{\lambda }_{2}}}}\left( {{\mathbf{p}}_{2}},{{\mathbf{k}}_{1}} \right)\hat{\gamma }_{{{\alpha }_{2}}}^{+}\left( {{\mathbf{p}}_{2}} \right){{{\hat{\gamma }}}_{{{\alpha }_{1}}}}\left( {{\mathbf{k}}_{1}}+\mathbf{p} \right){{{\hat{\gamma }}}_{{{\beta }_{2}}}}\left( {{\mathbf{p}}_{2}}-{{\mathbf{k}}_{1}} \right)}}\nonumber\\
	&+\frac{1}{2}\sum\limits_{\begin{smallmatrix} 
			{\lambda_{1}},{\alpha_{1}}\\ 
				{ \lambda_{2}},{\alpha_{2}}
		\end{smallmatrix}}{\sum\limits_{{{\mathbf{k}}_{2}},{{\mathbf{k}}_{1}}}{\Phi _{\alpha ,{{\alpha }_{1}},{{\lambda }_{1}}}^{*}\left( {{\mathbf{k}}_{1}}+\mathbf{p},{{\mathbf{k}}_{1}} \right)\Phi _{{{\alpha }_{1}},{{\alpha }_{2}},{{\lambda }_{2}}}^{*}\left( {{\mathbf{k}}_{1}}+\mathbf{p}+{{\mathbf{k}}_{2}},{{\mathbf{k}}_{2}} \right)\hat{\chi }_{{{\lambda }_{1}}}^{+}\left( {{\mathbf{k}}_{1}} \right)\hat{\chi }_{{{\lambda }_{2}}}^{+}\left( {{\mathbf{k}}_{2}} \right){{{\hat{\gamma }}}_{{{\alpha }_{2}}}}\left( {{\mathbf{k}}_{1}}+\mathbf{p}+{{\mathbf{k}}_{2}} \right)}}\nonumber\\
	&	+\dots.
		\label{eq32}
	\end{align}
	We do not give similar transformations for the operators $\hat{\chi }_{\lambda }^{+}\left( \mathbf{k} \right)$  and $\hat{\gamma }_{\alpha }^{+}\left( \mathbf{p} \right)$  here; they can be reproduced by the Hermitian conjugation operation of expressions \eqref{eq31}, \eqref{eq32}. Note that the ellipses at the end of formulas \eqref{eq31}, \eqref{eq32} mean the same as in formula \eqref{eq29}. Namely, they indicate that there may be two-operator and three-operator terms coming from commutators of a higher order than the second, for example, of type  $\frac{1}{3!}\left[ \left[ \left[ A,S \right],S \right],S \right]$. In addition to such terms, three-operator terms can also appear due to the interaction Hamiltonian  ${{\mathcal{\hat{V}}}_{2}}\left( \eta ,C \right)$, see \eqref{eq6}, \eqref{eq19}, if it is added to the truncated Hamiltonian in order to further diagonalize the obtained result. We do not write the terms of this type here, not because of fundamental difficulties with their calculation, but due to the cumbersome appearance of transformations \eqref{eq31}, \eqref{eq3} when they are taken into account. The questions whether such terms can be neglected or not are discussed below. In further formulas, we do not write such ellipses.

	\section{Influence of the photonic component on the dispersion law of quantum gases}
	
	Let us now apply the results obtained in the previous section (see \eqref{eq23}--\eqref{eq32}) to the diagonalization of the truncated Hamiltonian \eqref{eq25}. To do this, we substitute formulas \eqref{eq31}, \eqref{eq32} and formulas of Hermitian conjugates to them into \eqref{eq25}. In the obtained expressions, we should keep the two-operator and three-operator terms, neglecting the terms with a larger number of operators. The result of the ``assembly'' of the two-operator terms,  ${{\Sigma}_{2}}$, can be represented in the following form:
	\begin{equation}\label{eq33}
		{{\Sigma}_{2}}=\frac{1}{2}\sum\limits_{\alpha ,\beta }{{{\Sigma}_{\alpha ,\beta }}\left( \mathbf{p} \right)\hat{\gamma }_{\alpha }^{+}\left( \mathbf{p} \right){{{\hat{\gamma }}}_{\beta }}}\left( \mathbf{p} \right)+\mathrm{h.c.},
	\end{equation}
	where the value of ${{\mathrm{E} }_{\alpha ,\beta }}\left( \mathbf{p} \right)$  is given by expression
	\begin{align}
		{{\Sigma }_{\alpha ,\beta }}=\frac{1}{2}{{\varepsilon }_{\alpha }}\left( \mathbf{p} \right){{\delta }_{\alpha ,\beta }}&+\sum\limits_{\gamma ,\lambda ,\mathbf{k}}{\left\{ \left[ {{\varepsilon }_{\gamma }}\left( \mathbf{p}-\mathbf{k} \right)+{{\omega }_{\mathbf{k}}}-{{\varepsilon }_{\alpha }}\left( \mathbf{p} \right) \right]{{\Phi }_{\alpha ,\gamma ,\lambda }}\left( \mathbf{p},\mathbf{k} \right)\Phi _{\gamma ,\beta ,\lambda }^{*}\left( \mathbf{p},\mathbf{k} \right) \right.}\nonumber\\
	&	+\left[ {{\varepsilon }_{\gamma }}\left( \mathbf{p}-\mathbf{k} \right)+{{\omega }_{\mathbf{k}}}-{{\varepsilon }_{\beta }}\left( \mathbf{p} \right) \right]{{\Phi }_{\beta ,\gamma ,\lambda }}\left( \mathbf{p},\mathbf{k} \right)\Phi _{\gamma ,\alpha ,\lambda }^{*}\left( \mathbf{p},\mathbf{k} \right)\nonumber\\
	&	-\left. {{F}_{\alpha ,\gamma ,\lambda }}\left( \mathbf{p},\mathbf{k} \right)\Phi _{\gamma ,\beta ,\lambda }^{*}\left( \mathbf{p},\mathbf{k} \right) - {{\Phi}_{\beta ,\gamma ,\lambda }}\left( \mathbf{p},\mathbf{k} \right) F _{\gamma ,\alpha ,\lambda }^{*}\left( \mathbf{p},\mathbf{k} \right)\right\}.
		\label{eq34}
	\end{align}
	%
	We recall that the value ${{F}_{\alpha ,\beta ,\lambda }}\left( \mathbf{p},\mathbf{k} \right)$   defines the operator  ${{\mathcal{\hat{V}}}_{1}}\left( \eta ,C \right)$, see \eqref{eq13}, and ${{\Phi }_{\alpha ,\gamma ,\lambda }}\left( \mathbf{p},\mathbf{k} \right)$  is the operator of canonical transformations \eqref{eq24}.
	
	The result of the assembly of three-operator terms,  ${{\Sigma}_{3}}$, can be represented by the following expression:
	\begin{align}
	{{\Sigma}_{3}}&=\sum\limits_{\alpha ,\beta ,\lambda }{\sum\limits_{\mathbf{p},\mathbf{k}}{\left\{ {{F}_{\alpha ,\beta ,\lambda }}\left( \mathbf{p},\mathbf{k} \right)+{{\Phi }_{\alpha ,\beta ,\lambda }}\left( \mathbf{p},\mathbf{k} \right)\left[ {{\varepsilon }_{\alpha }}\left( \mathbf{p} \right)-{{\omega }_{\mathbf{k}}}-{{\varepsilon }_{\beta }}\left( \mathbf{p}-\mathbf{k} \right) \right] \right\}}} \nonumber\\
	&\times \hat{\gamma }_{\alpha }^{+}\left( \mathbf{p} \right){{\hat{\gamma }}_{\beta }}\left( \mathbf{p}-\mathbf{k} \right){{\hat{\chi }}_{\lambda }}\left( \mathbf{k} \right)+\mathrm{h.c.}
	\label{eq35}
\end{align}
	We have already noted that canonical transformations are introduced in order to get rid of three-operator terms. As seen from \eqref{eq35}, this can be done by equating to zero the expression in the braces, which allows us to find the value  ${{F}_{\alpha ,\beta ,\lambda }}\left( \mathbf{p},\mathbf{k} \right)$:
	\begin{equation}\label{eq36}
		{{\Phi }_{\alpha ,\beta ,\lambda }}\left( \mathbf{p},\mathbf{k} \right)=\frac{{{F}_{\alpha ,\beta ,\lambda }}\left( \mathbf{p},\mathbf{k} \right)}{{{\varepsilon }_{\beta }}\left( \mathbf{p}-\mathbf{k} \right)+{{\omega }_{\mathbf{k}}}-{{\varepsilon }_{\alpha }}\left( \mathbf{p} \right)}.
	\end{equation}
	Expression \eqref{eq36}, in turn, makes it possible to calculate the tensor function of the momentum  ${{\mathrm{E} }_{\alpha ,\beta }}\left( \mathbf{p} \right)$, using expression \eqref{eq33}:
\begin{align}
&	{{\Sigma}_{\alpha ,\beta }}\left( {\mathbf{p}} \right) =\frac{1}{2} {\varepsilon _\alpha }\left( {\mathbf{p}} \right){\delta _{\alpha ,\beta }}  \nonumber\\
&	- \sum\limits_{\gamma ,\lambda ,{\mathbf{k}}} {{F_{\beta ,\gamma ,\lambda }}\left( {{\mathbf{p}},{\mathbf{k}}} \right)F_{\gamma ,\alpha ,\lambda }^*\left( {{\mathbf{p}},{\mathbf{k}}} \right)\left\{ {\frac{1}{{{\varepsilon _\alpha }\left( {{\mathbf{p}} - {\mathbf{k}}} \right) + {\omega _{\mathbf{k}}} - {\varepsilon _\gamma }\left( {\mathbf{p}} \right)}} - \frac{1}{{{\varepsilon _\gamma }\left( {{\mathbf{p}} - {\mathbf{k}}} \right) + {\omega _{\mathbf{k}}} - {\varepsilon _\beta }\left( {\mathbf{p}} \right)}}} \right\}},	
	\label{eq37}
\end{align}
	where ${{F}_{\alpha ,\beta ,\lambda }}\left( \mathbf{p},\mathbf{k} \right)$  is determined by formulas \eqref{eq14}--\eqref{eq18}.
	
	Now, the truncated Hamiltonian ${{\mathcal{\hat{H}}}_{\mathrm{tr}}}\left( \gamma ,\chi  \right)$  takes the form
	\begin{equation}\label{eq38}
		{{\mathcal{\hat{H}}}_{\mathrm{tr}}}\left( \gamma ,\chi  \right)=\sum\limits_{\alpha ,\beta }{{{\mathrm{E} }_{\alpha ,\beta }}\left( \mathbf{p} \right)\hat{\gamma }_{\alpha }^{+}\left( \mathbf{p} \right){{{\hat{\gamma }}}_{\beta }}\left( \mathbf{p} \right)}+\sum\limits_{\mathbf{k},\lambda }{\omega \left( \mathbf{k} \right)\hat{\chi }_{\lambda }^{+}\left( \mathbf{k} \right){{{\hat{\chi }}}_{\lambda }}}\left( \mathbf{k} \right),
	\end{equation}
	\[
	{{\mathrm{E} }_{\alpha ,\beta }}=\Sigma_{\alpha ,\beta }(\mathbf{p})+\Sigma_{\beta,\alpha }^{*}(\mathbf{p}).
	\]
	As we can see, the ``new'' dispersion law of atoms of quantum gases ${{\mathrm{E} }_{\alpha ,\beta }}\left( \mathbf{p} \right)$  due to interaction with radiation (photons) is significantly deformed in comparison with  ${{\varepsilon }_{\alpha }}\left( \mathbf{p} \right)$, see \eqref{eq4}, stopping to be diagonal over the quantum mechanical characteristics of states. The dispersion law of photons  $\omega \left( \mathbf{k} \right)$, due to the application of canonical transformations defined by formulas \eqref{eq23}--\eqref{eq32}, remains undeformed. We should recall here that a similar situation is observed in the case of redefining the dispersion laws of electrons due to interaction with phonons in a solid, as shown in \cite{16}. In this paper, the situation is somewhat different, since the system is characterized by an additional Hamiltonian  ${{\mathcal{\hat{V}}}_{2}}\left( \eta ,C \right)$, see \eqref{eq6}, \eqref{eq19}, also associated with the interaction of atoms and photons. There is no such an interaction operator in \cite{16}. In our opinion, the mentioned Hamiltonian can be responsible for reformatting the spectrum of photons in matter. However, this issue requires additional research.
	
	Note that in the expression for the dispersion law \eqref{eq37}, it is possible to perform summation over polarizations $\lambda $  if we use the explicit form of the value  ${{F}_{\alpha ,\beta ,\lambda }}\left( \mathbf{p},\mathbf{k} \right)$, see \eqref{eq14}--\eqref{eq18}, and take into account the validity of the relation for polarization vectors: 
	\begin{equation}\label{eq39}
		\sum\limits_{\lambda }{{{\left( \mathbf{e}_{\mathbf{k}}^{\left( \lambda  \right)} \right)}_{i}}{{\left( \mathbf{e}_{\mathbf{k}}^{\left( \lambda  \right)} \right)}_{j}}}={{\delta }_{i,j}}-\frac{{{k}_{i}}{{k}_{j}}}{{{k}^{2}}}.
	\end{equation}
	As a consequence of calculating the sum over  $\lambda$, the dispersion law of atoms of quantum gases interacting with photons can be written as:
	\begin{align}
	&	{{\mathrm{E} }_{\alpha ,\beta }}\left( \mathbf{p} \right)={{\varepsilon }_{\alpha }}\left( \mathbf{p} \right){{\delta }_{\alpha ,\beta }}\nonumber\\
		& + \frac{{2\piup }}{{{c^2}V}}\sum\limits_{\gamma ,{\bf{k}}} {\left\{ {{{\bf{J}}_{\alpha ,\gamma }}\left( {{\bf{p}},{\bf{k}}} \right){\bf{J}}_{\gamma ,\beta }^*\left( {{\bf{p}},{\bf{k}}} \right) - \frac{{\left( {{\bf{k}}{{\bf{J}}_{\alpha ,\gamma }}\left( {{\bf{p}},{\bf{k}}} \right)} \right)\left( {{\bf{kJ}}_{\gamma ,\beta }^*\left( {{\bf{p}},{\bf{k}}} \right)} \right)}}{{{k^2}}}} \right\}}  \nonumber\\
		& \times \left\{ {\frac{1}{{ {{\varepsilon _\beta }\left( {{\bf{p}} - {\bf{k}}} \right) + {\omega _{\bf{k}}} - {\varepsilon _\gamma }\left( {\bf{p}} \right)} }} - \frac{1}{{{\varepsilon _\gamma }\left( {{\bf{p}} - {\bf{k}}} \right) + {\omega _{\bf{k}}} - {\varepsilon _\alpha }\left( {\bf{p}} \right)}}} \right\} \nonumber\\
		& + \frac{{2\piup }}{{{c^2}V}}\sum\limits_{\gamma ,{\bf{k}}} {\left\{ {{{\bf{J}}_{\beta ,\gamma }}\left( {{\bf{p}},{\bf{k}}} \right){\bf{J}}_{\gamma ,\alpha }^*\left( {{\bf{p}},{\bf{k}}} \right) - \frac{{\left( {{\bf{k}}{{\bf{J}}_{\beta ,\gamma }}\left( {{\bf{p}},{\bf{k}}} \right)} \right)\left( {{\bf{kJ}}_{\gamma ,\alpha }^*\left( {{\bf{p}},{\bf{k}}} \right)} \right)}}{{{k^2}}}} \right\}}  \nonumber\\
		& \times\left\{ {\frac{1}{{{\varepsilon _\alpha }\left( {{\bf{p}} - {\bf{k}}} \right) + {\omega _{\bf{k}}} - {\varepsilon _\gamma }\left( {\bf{p}} \right)}} - \frac{1}{{{\varepsilon _\gamma }\left( {{\bf{p}} - {\bf{k}}} \right) + {\omega _{\bf{k}}} - {\varepsilon _\beta }\left( {\bf{p}} \right)}}} \right\}.
		\label{eq40}
	\end{align}
	We can also hope to further simplify the dispersion law if we use in \eqref{eq40} the expression \eqref{eq17} ${{\mathbf{J}}_{\alpha ,\beta }}\left( \mathbf{p},\mathbf{k} \right)$  in the point approximation for an atom. But again, this approximation requires additional research into the possibility of its application, since the mentioned simplification of \eqref{eq17} in \eqref{eq40} may cause problems with calculating the sum over  $\mathbf{k}$.
	
	However, we believe that the main task announced in this article, namely, the demonstration of the prospects of the method of canonical transformations in the theory of quantum gases interacting with radiation, can be considered solved. The evidence of this statement is the expressions \eqref{eq36}--\eqref{eq40} for the dispersion laws of atoms ``disguised'' due to the specified interaction. According to the material presented above, the movement of atoms in the ``coat'' of virtual photons should be considered as the reasons for such a disguise. As a result, such atoms, which become quasiparticles, can be ``assigned'' an effective mass.
	 
	As is known, in the theory of a solid body, in the general case, the effective mass of an electron is a tensor quantity (see, for example, \cite{17, 18} and the literature cited therein regarding this subject), which is defined as follows:
	\begin{equation}\label{eq41}
		{{\hbar }^{2}}m_{ij}^{*-1}={{\left( \frac{{{\partial }^{2}}E\left( \mathbf{k} \right)}{\partial {{k}_{i}}\partial {{k}_{j}}} \right)}_{\mathbf{k}={{\mathbf{k}}_{0}}}},
	\end{equation}
	where $E\left( \mathbf{k} \right)$   is the anisotropic dispersion law of a dressed electron, ${{\mathbf{k}}_{0}}$  is the Brillouin zone point at which the dispersion curve reaches a minimum. Hence, it is clear that in this case \eqref{eq41} we are speaking about the  anisotropy by momentum of the dispersion law of a dressed electron in a solid.
	
	In case of \eqref{eq40} we are dealing, along with the anisotropy by momentum, with the polarization of the dispersion law ${{\mathrm{E} }_{\alpha ,\beta }}\left( \mathbf{p} \right)$   of the dressed atom in the sets of quantum numbers  $\alpha$, $\beta$, see \eqref{eq15}, \eqref{eq17}. That is, if we proceed by analogy with \eqref{eq41}, we can define the effective mass tensor as follows:
	\begin{equation}\label{eq42}
		m_{i,j;\alpha ,\beta }^{*-1}={{\left( \frac{{{\partial }^{2}}{{\mathrm{E} }_{\alpha ,\beta }}\left( \mathbf{p} \right)}{\partial {{p}_{i}}\partial {{p}_{j}}} \right)}_{\mathbf{p}={{\mathbf{p}}_{0}}\left( \alpha ,\beta  \right)}},
	\end{equation}
	from which it follows that the effective mass is a tensor not only in the space of momentum, but also has a tensor structure in the indices  $\alpha$, $\beta$, which characterize the quantum mechanical states of the undressed atom. That is, the effective mass in this case depends on specific pairs of the set of quantum characteristics of the energy levels   $\alpha$, $\beta$ of the atom, on which the momentum  ${{\mathbf{p}}_{0}}\left( \alpha ,\beta  \right)$ should also depend. According to \eqref{eq42}, this momentum should determine the point of extremum of the value  ${{\mathrm{E} }_{\alpha ,\beta }}\left( \mathbf{p} \right)$. In other words, the effective mass of the atom depends on the energy levels $\alpha$, $\beta$  of the atom which are connected by a cloud of virtual photons. In the case of the effective mass of an electron in a solid, the procedures are established for diagonalizing the effective mass tensor  $m_{ij}^{*-1}$, see \eqref{eq41}. As for the case \eqref{eq42} of this paper, the anisotropy of the value  ${{\mathrm{E} }_{\alpha ,\beta }}\left( \mathbf{p} \right)$  by momentum in this expression should disappear after performing summation over $\mathbf{k}$  in \eqref{eq40}, and the effective mass will become proportional to the Kronecker symbol  ${{\delta }_{ij}}$. The validity of such a statement follows from general considerations. Indeed, this should happen because in the problem we have formulated there are no directions along which such anisotropy could be formed. As for the tensor structure of the dispersion law by indices  $\alpha$, $\beta$, here one may think about further diagonalization of the value  ${{\mathrm{E} }_{\alpha ,\beta }}\left( \mathbf{p} \right)$. However, several difficult questions immediately arise regarding this. Questions related both to the very possibility of such diagonalization and to the correct introduction of the corresponding procedure. Currently, such questions are under study.

	Finally, we return to discussing the issues we announced in the end of section~\ref{Sec2}. These issues concern the analysis of the presence in expressions \eqref{eq28}--\eqref{eq32} of the terms related to the inclusion in \eqref{eq29} of the terms of the type
	\[
	+\frac{1}{3!}\left[ \left[ \left[ A,S \right],S \right],S \right]+\frac{1}{4!}\left[ \left[ \left[ \left[ A,S \right],S \right],S \right],S \right]+\dots
	\,.\]
	Taking into account such two terms leads to the induction in expressions \eqref{eq31}, \eqref{eq32}, and hence \eqref{eq33}--\eqref{eq35} of the terms having the third and fourth order in value  ${{\Phi }_{\alpha ,\beta ,\lambda }}\left( \mathbf{p},\mathbf{k} \right)$, which determines the canonical transformation \eqref{eq24}. As a result, the expression with three-operator terms will have a much more complex structure than \eqref{eq35}. It will include the value ${{\mathrm{E} }_{\alpha ,\beta }}\left( \mathbf{p} \right)$  itself, which, in turn, contains a nonlinear dependence on  ${{\Phi }_{\alpha ,\beta ,\lambda }}\left( \mathbf{p},\mathbf{k} \right)$. Thus, by setting the three-operator terms to zero, a complex nonlinear equation for tensor values is obtained. It should be noted that in \cite{16} they also come to a similar problem, with the difference that therein the complex nonlinear equations are not obtained for tensor values. In \cite{16}, such an equation is proposed to be solved by the method of iterations. If we also solve by iterations the nonlinear equations for  ${{\Phi }_{\alpha ,\beta ,\lambda }}\left( \mathbf{p},\mathbf{k} \right)$  in our case, then the first iteration will have the form~\eqref{eq36}, and therefore the dispersion law ${{\mathrm{E} }_{\alpha ,\beta }}\left( \mathbf{p} \right)$   will be determined by formula \eqref{eq40}. As can be seen from this formula, the need for subsequent iterations may be justified under circumstances when peculiarities may arise in the denominator of expression \eqref{eq40}. However, such a problem should be considered only in the case of concretization of a physical system, for example, by determination with an element whose atoms are considered non-hydrogen-like, atomic levels of a specific non-hydrogen-like atom, etc. In this article, we do not consider such a concretization.

\section*{Acknowledgements}

	M.~S.~Bulakhov acknowledges support by the National Research Foundation of Ukraine, Grant No.~0124U004372.
	A.~S.~Peletminskii and Yu.~V.~Slyusarenko acknowledge the STCU project ``Magnetism in Ukraine Initiative'', Grant No.~9918.

	\bibliographystyle{cmpj}

	%
	%
\newpage

\ukrainianpart

\title{Метод канонічних перетворень в теорії квантових газів, взаємодiючих із випромінюванням}
\author{М. С. Булахов\refaddr{label1},
	О. С. Пелетминський\refaddr{label1},
	П. П. Костробій\refaddr{label2},
	І. А. Рижа\refaddr{label2},
	Ю.~В.~Слюсаренко\refaddr{label1,label2,label3}}
\addresses{
	\addr{label1}  Інститут теоретичної фізики ім.~О.~І.~Ахієзера ННЦ ХФТІ, вул.~Академічна, 1, 61108 Харків, Україна
	\addr{label2} Національний університет ``Львіська політехніка'', вул.~С.~Бандери, 12, 79013 Львів, Україна
	\addr{label3} Харківський національний університет ім. В.~Н.~Каразіна, пл.~Свободи, 4,  61022 Харків, Україна
}
%
%
%

\makeukrtitle

\begin{abstract}
	\tolerance=3000%
	Запропоновано підхід до теоретичного дослідження ефектів та явищ у квантових газах, взаємодiючих із випромінюванням. Підхід базується на модифікації методу канонічних перетворень, який свого часу було застосовано для діагоналізації  гамільтоніанів, що описують взаємодію електронів із фононами в твердому тілі. Можливості методу продемонстровано на вивченні впливу фотонів на спектральні характеристики атомів квантових газів, взаємодiючих із випромінюванням. У рамках розвинутого підходу досліджено ефект ``одягання'' атомів квантових газів за рахунок хмари віртуальних фотонів і здобуто вирази для енергетичних характеристик таких перевдягнутих атомів --- квазічастинок. Обговорено проблему визначення поняття ефективної маси таких квазічастинок.
	\keywords квантові гази, фотони, канонічні перетворення, одягнуті атоми, закон дисперсії, ефективна маса
	
\end{abstract}

\end{document}